\documentclass[event]{sigirforum}
\def\pubissue{Vol. 58 No. 2 -- December 2024}

\def\eventdate{18 July 2024}
\def\eventurl{https://sim4ia.org/sigir2024}

\usepackage{booktabs}
\usepackage{tabularx}
\usepackage{pbox}

\usepackage[table, x11names]{xcolor}
\usepackage{array, booktabs, boldline} %
\usepackage{cellspace}

\usepackage{dirtytalk}
\begin{document}
% USE TITLE CASE FOR THE TITLE

\title{Report on the Workshop on Simulations for Information Access (Sim4IA 2024) at SIGIR~2024}

\authors{
    %\author[mail]{Name}{University}{City, Country}
    \author[timobreuer@acm.org]{Timo Breuer}{TH Köln}{Cologne, Germany}
    \and
    \author[ckreutz@acm.org]{Christin Katharina Kreutz}{TH Mittelhessen}{Gießen, Germany}
    \and
    \author[norbert.fuhr@uni-due.de]{Norbert Fuhr}{University of Duisburg-Essen}{Duisburg, Germany}
    \and
    \author[krisztian.balog@uis.no]{Krisztian Balog}{University of Stavanger}{Stavanger, Norway}
    \and
    \author[philipp.schaer@th-koeln.de]{Philipp Schaer}{TH Köln}{Cologne, Germany}
    \and
    %
    % please insert your name + affiliation here 
    %
  \\
  % alphabetical order
\parbox{\columnwidth}{
\centering
Nolwenn Bernard, Ingo Frommholz, Marcel Gohsen, Kaixin Ji, Gareth J. F. Jones, Jüri Keller, Jiqun Liu, Martin Mladenov, Gabriella Pasi, Johanne Trippas, Xi Wang, Saber Zerhoudi, ChengXiang Zhai%\footnote{Affiliation not shown for all authors due to space limitations (see Appendix~\ref{sec:appendix} for details).}
}
}

\maketitle 
\begin{abstract}
This paper is a report of the Workshop on Simulations for Information Access (Sim4IA) workshop at SIGIR 2024. The
workshop had two keynotes, a panel discussion, nine lightning talks, and two breakout sessions. Key takeaways were user simulation's importance in academia and industry, the possible bridging of online and offline evaluation, and the issues of organizing a companion shared task around user simulations for information access. 
We report on how we organized the workshop, provide a brief overview of what happened at the workshop, and summarize the main topics and findings of the workshop and future work.

\end{abstract}

\section{Introduction}

The common approach and general understanding of evaluating information access systems (like search engines, recommender systems, or conversational agents) is closely coupled to the Cranfield paradigm, the dominating evaluation method, especially in information retrieval (IR). This has proven to be able to deal with the inherent complexity in information access contexts. The Cranfield studies can be understood to use a special form of simulation to mimic the search process by making implicit and explicit assumptions about the information system and its users. This helps to reduce the complexity of the search process and allows us to effectively compare different IR systems. Despite its long history and roots within the community, Cranfield has not been without criticism \citep{ingwersen_turn_2005} and the underlying assumptions are often described as (over-)simplifications leading to potentially unrealistic search evaluations that deviate from users' actual interaction experience and search task performance \citep{chen2023reference}.

Other evaluation methods, including interactive/session-based retrieval settings and controlled user experiments~\citep{kelly_methods_2009, liu2019interactive}, living labs~\citep{hopfgartner_continuous_2019}, or (user) simulation studies~\citep{balog2024user} have been proposed and discussed in the community; these have also been used in shared tasks at TREC, NTCIR, or CLEF, e.g. iCLEF~\citep{DBLP:conf/clef/GonzaloPCK09}, OpenSearch~\citep{jagerman-2017-overview}, and LiLaS \citep{schaer-2021-overview}). However, no shared tasks at TREC/CLEF have primarily focused on user simulations. %Very few labs were concerned with user interactions and their simulation. One example is NewsREEL \citep{DBLP:conf/clef/KilleLGHLSMSBV16} where so-called replay data was recorded in the online evaluation of the lab to simulate user interactions. 
Recently, in the TREC Interactive Knowledge Assistance Track (iKAT)~\citep{aliannejadi2024trec}, some submissions included simulated user feedback in their interactive information access systems, while the lab did not employ such an evaluation strategy.
%These alternative evaluation endeavors aim to enable a more realistic and holistic information access evaluation by including richer user models or more complex representations of the search processes (like sessions). 
Simulations can also contribute to a better understanding of users. Formalizing a user model for simulation delivers explicit hypotheses on user behavior, which can produce insights into the validity of assumptions about users~\citep{balog2024user}. 

Other recent examples of a re-started interest in the topic of (user) simulation were the Sim4IR workshop that was held at SIGIR 2021~\citep{DBLP:journals/sigir/BalogMTZ21}, the SIMIIR 2.0 framework\footnote{\url{https://github.com/padre-lab-eu/simiir-2}}~\citep{10.1145/3511808.3557711}, tutorials~\citep{DBLP:conf/cikm/BalogZ23}, and a recurring theme of how generative model can be used for simulation~\citep{azzopardi2024report}.
At ECIR or SIGIR, a reasonable number of relevant papers on user simulations were accepted, and even a study on simulating user queries won the best paper award at ECIR 2022~\citep{DBLP:conf/ecir/PenhaCH22}. Additionally, the introduction of generative AI methods opened up new possibilities for integrating LLMs to simulate users. 

Therefore, to understand how and whether the evaluation of information access technology can truly benefit from simulating user interactions, we organized the first
workshop on Simulations for Information Access (Sim4IA 2024), held in conjunction with SIGIR 2024. Its aim was to serve as a forum to bring together researchers and experts.
Additionally, this workshop's goal was to provide a much-needed forum for the community to discuss the emerging challenges when applying (user) simulations to evaluate information access systems in simulation-based shared tasks.

%, building on past efforts to create ``a forum for researchers and practitioners to promote methodology and development of more widespread use of simulation for IR evaluation.'' 
%Around 80 participants demonstrated the interest in the topic and the engagement from the community.
%The main conclusions from the workshop's discussion were ``that simulation has the potential to offer solutions to the limitations of existing evaluation methodologies, but there is more research needed toward developing realistic user simulators; and the development and sharing of simulators, in the form of toolkits and online services, is critical for successful uptake''~\citep{DBLP:journals/sigir/BalogMTZ21}.

%In this current workshop\footnote{\url{https://sim4ia.org/sigir2024/}} we dwell deeper into the field by focusing on the user simulation part. This form of simulation recently gained popularity due to available toolkits like SIMIIR 2.0\footnote{\url{https://github.com/padre-lab-eu/simiir-2}}~\cite{10.1145/3511808.3557711} or tutorials on this exact topic (CIKM 2023~\cite{DBLP:conf/cikm/BalogZ23}). 

%Nevertheless, a specific venue to present and discuss new and experimental approaches is missing. No track or workshop on simulations was present at ECIR or SIGIR 2022/2023. However, a reasonable number of relevant papers on user simulations were accepted, and even a study on simulating user queries won the best paper award at ECIR 2022 \citep{DBLP:conf/ecir/PenhaCH22}. Additionally, since the introduction of generative AI methods into the field, the possibility of integrating LLMs to simulate users has opened up a new chapter. 

This paper is a report of the Sim4IA\footnote{\url{https://sim4ia.org/sigir2024/}}~\citep{Schaer2024} workshop at SIGIR 2024. The workshop had two keynotes, a panel discussion, nine lightning talks, and two breakout sessions. 
We report on how we organized the workshop, provide a brief overview of what happened at the workshop, and summarise the main topics and findings of the workshop as well as future work.

\section{Workshop Overview}

Sim4IA was a full-day workshop at SIGIR 2024, held in Washington, D.C., on 18 July 2024. The workshop attracted 25 participants who participated in a very interactive setting. Instead of a typical ``mini-conference'' we decided to focus on short, but thought-provoking lightning talks from the participants, two keynotes and a panel discussion (see Table~\ref{tab:schedule}). Participants could later join two breakout discussion groups to deepen the previous discussions and further outline future research topics and methods. 

To enable interaction with a broader set of participants, we offered limited hybrid participation in addition to onsite attendance via Zoom and Slack. 

\begin{table}[t]
    \centering
    \begin{tabular}{ll}
    \toprule
    Time & Agenda\\ \midrule
        9:00--9:15 & Welcome \\
        9:15--10:00 & Keynote 1: Gabriella Pasi \\
        10:00--10:30 & Lightning talks, talks 1 - 4 (5 minutes each) \\
        \rowcolor{gray!30!} 10:30--11:00 & Coffee break \\
        11:00--12:00 & Panel discussion \\ 
        12:00--12:30 & Lightning talks, talks 5 - 9 (5 minutes each) \\
        \rowcolor{gray!30!} 12:30--13:30 & Lunch break\\
        13:30--14:15 & Keynote 2: Martin Mladenov\\
        14:15--15:00 & Breakout group discussions I \\
        \rowcolor{gray!30!} 15:00--15:30 & Coffee break\\
        15:30--16:15 & Breakout group discussions II\\
        16:15--17:00 & Reports of the group discussions and closing \\ \bottomrule
    \end{tabular}
    \caption{Timeline of the Sim4IA workshop.}
    \label{tab:schedule}    
\end{table}

\section{Keynotes}

Our keynote speakers, Gabriella Pasi (University of Milano Bicocca) and Martin Mladenov (Google), both delivered their keynotes before taking questions from the audience. They represented perspectives from academia and industry. 

Gabriella Pasi's keynote addressed the issue of personalizing information access by leveraging the user's experience, preferences and expertise. In particular, personalized search has been a core research focus for many years to offer users a search experience that can improve accessibility to content that is retrieved in response to their queries. This task involves two primary sub-tasks: modeling users and their context, and leveraging user models to constrain the search process towards producing a personalized outcome. Personalization can be interpreted as a simulation process, where the system relies on ``knowledge'' about a user to select content that is possibly useful and accessible to the specific user. Seen through this lens, effective and correct user modeling is paramount to an effective user simulation. In this perspective, the talk raised some key questions about the two above aspects.

Martin Mladenov's keynote focused on the application of user simulation as an engineering tool. Results at Google indicate that calibrated user simulations show promise to replace (at least partially) A/B tests as the core driver of the recommender system development cycle. The keynote emphasized that before this promise can be fulfilled, and user simulation could become a standard part of the recommender system development toolkit, a number of open questions need to be understood. These questions revolve around applicability, credibility, and reliability. The talk outlined potential approaches towards answering these questions in terms of developing diagnostics for individual simulation models as well as theoretical guarantees for the general simulation-driven development process. The talk introduced RecSim NG as a tool for developing solutions.

\section{Panel Discussion}

Besides the keynotes and lightning talks, the workshop featured a panel discussion with four invited panelists, including the two keynote speakers, Gabriella Pasi and Martin Mladenov, Johanne Trippas (RMIT University), and ChengXiang Zhai (University of Illinois at Urbana-Champaign). The panelists shared their experiences, opinions, and stances on simulations. They were moderated by Norbert Fuhr, who organized the discussion around the following three questions (see the left side of \autoref{fig:panel_and_breakout}).

\subsection*{What is the purpose of simulating users?}

Overall, the panelists mainly agreed that it is quite challenging to pinpoint a single purpose of user simulations, as the use cases and benefits for real users are manifold. Among many terrific ideas as to why user simulations can be useful, the following aspects were highlighted by the panelists. Personalization can be understood as a simulation process. In this sense, simulations can help users to better access useful (and personalized) information. Besides a deeper understanding of the user, simulations also help enable better understanding of the system and make the engineering process more transparent. They are particularly useful for evaluating an interactive system and making evaluations reproducible. Usually, the user's knowledge state changes as the session progresses and after the experiment is finished. In this regard, simulating users allows better understanding at different stages of the search process with explicit modeling of stage transitions. Enabling interpretability is another important merit factor, as simulations are grounded on a testable user model. Set against the prevailing agreement on the usefulness of user simulations, panelists also highlighted their limitations. If the underlying user model is incorrect, the simulations would not make much sense. Furthermore, in some cases, user simulations can imply an abstraction that goes too far to allow any generalizable conclusions.

\subsection*{For what kinds of experiments and evaluations have you used user simulation?} % did you use them before?}

From a more personal point of view, the panelists shared their experiences with applying user simulations in the experimentation and evaluation process. First and foremost, user simulations enable offline experimentation without involving real users in the early development cycles. Doing so allows testing a system without conducting a sometimes risky and expensive A/B test. Sharing personal experiences and anecdotes, one panelist reported that simulators are sometimes more accurate than A/B tests, as they better align with metrics from the production systems, leaving opportunities for interesting research questions about why this is the case. Often, the simulators are based on real user logs, although there are limitations on how far they can be used for insightful estimates. It can be quite challenging to make reliable estimates for an out-of-distribution problem setting, where logs are obtained from a possibly different population or environment to estimate a new system feature, for example.

More user and context data is particularly helpful for reliable user models. In this regard, the academic search setting offers a profound basis for obtaining this kind of data from publications, as outlined by one of the panelists. For instance, a user's knowledge state can be modeled based on the cited works of a publication, helping to generate data where it is usually unavailable. Another panelist emphasized the usefulness of simulators in evaluating interface alternatives. For instance, two or more interfaces can be compared for a known-item search with regard to how much effort is required to reach the known item in the session. Even though simulations can be imperfect, they often allow a reliable evaluation of which systems are better than a baseline or at least how they differ.

Last but not least, one panelist also shared experiences with user simulations as useful tools for robustness tests of production systems. Even simple user models are often enough to run security checks or test the trustworthiness of a platform by spoiling the user base with fake users. As part of the discussions with the audience, the panelists also discussed the idea of having a guiding system that helps users during a search session by predicting the next interaction steps with a reasonable user model. Everyone agreed that such a system would be particularly helpful in the context of a conversational system.

\subsection*{How realistic are our user simulations?}

One skeptical panelist argued that relying on (unrealistic) user simulations can be misleading and that they have to be critically analyzed. As an example, the panelist was referring to the field of aerospace engineering, where turbulences are an ongoing subject of simulations. Even though air travel is considered to be safe when people have buckled up, injuries or even deaths occur because people simply do not use their belts. If the simulation does not cover these cases, relying on simplistic models can be harmful in the extreme case. 

Nevertheless, it was also argued that simulations are an important tool for better risk estimations, as they mainly help reduce entropy and uncertainty. Still, it was also pointed out that modeling the entire user might be too complex. A user model always implies a certain kind of abstraction, and not every aspect of the user behavior has to be covered by the model as long as it satisfies the requirements of the experimental setup and is sufficient to answer the underlying research question. For instance, sometimes, it is sufficient to have rather simple user simulators that are good enough to distinguish between two systems for which the effectiveness is known a priori.

In this regard, simulated user interactions must be analyzed carefully, especially the generalizability of the conclusions that can be drawn from them. Very often, user models focus on particular aspects of the user behavior, which are usually related to specific tasks. For better generalizability and a more comprehensive approach to user modeling,
our community probably needs the help of others, e.g. psychology, as argued by one panelist, and better characterize and represent users' bounded rationality, interaction intents, and judgment strategies in search sessions. All panelists agreed that these cross-disciplinary approaches toward user simulations can be fostered by collaborations between industry and academia, and researchers from different disciplines. Most notably, academic researchers have a strong interest in obtaining data from real-world experiments, whereas participants with an industrial background mentioned that many models from academia help design experiments and products. Participants from academia and industry had a strong interest and willingness to bring these kinds of collaborations forward to advance the fidelity of user simulations.

\section{Lightning Talks}

A total of nine lightning talks were given, spread out over two designated sessions. Three of these were given remotely via Zoom. The time frame for each lighting talk was 5 minutes. Table~\ref{tab:lighting} summarizes all nine talks and shows the wide range of different topics covered in these talks. Six out of the presentations were re-submission of previously presented work: \citet{10.1007/978-3-031-56060-6_25}, \citet{sadiri-javadi-etal-2023-opinionconv}, \citet{10.1007/978-3-031-56027-9_10}, \citet{DBLP:conf/sigir/ZhangWGLM24}, \citet{10.1145/3626772.3661358}, \citet{10.1145/3664190.3672529}. 
% Please add your publications here, if you would like to include it in the paper.
The rest of the talks were original content. 

\begin{table}[t]
    \centering
    \begin{tabularx}{\linewidth}{X X}
    \toprule
    Authors & Title \\ \midrule
    \textbf{Saber Zerhoudi} and Michael Granitzer & DuSS: Exploring the Synergy Between Conversational Search and Traditional SERPs in Information Retrieval\\
    Johannes Kiesel, \textbf{Marcel Gohsen}, Nailia Mirzakhmedova, Matthias Hagen, and Benno Stein & Simulating Follow-up Questions in Conversational Search (remote talk) \\
    \textbf{Vahid Sadiri Javadi} and Lucie Flek & OpinionConv: A Framework for Simulating Opinionated Conversations for Product Search (remote talk)\\
    \textbf{Xi Wang}, Procheta Sen, Ruizhe Li, and Emine Yilmaz & Enhancing Conversational Techniques: The Role of Synthetic Dialogue Generation (remote talk)\\
    \textbf{Saber Zerhoudi} and Michael Granitzer & Beyond Conventional Metrics: Assessing User Simulators in Information Retrieval\\
    \textbf{Jüri Keller}, Björn Engelmann, Christin Kreutz, and Philipp Schaer &	Towards Information Nugget-Based Test Collections for Evaluating Information Access Systems\\
    \textbf{Erhan Zhang}, Xingzhu Wang, Peiyuan Gong, Yankai Lin, and Jiaxin Mao &	USimAgent: Large Language Models for Simulating Search Users\\
    \textbf{Chih-Wei Hsu}, Martin Mladenov, Ofer Meshi, James Pine, Hubert Pham, Shane Li, Xujian Liang, Anton Polishko, Li Yang, Ben Scheetz, and Craig Boutilier & Minimizing Live Experiments in Recommender Systems: User Simulation to Evaluate Preference Elicitation Policies\\
    \textbf{Nolwenn Bernard} and Krisztian Balog & Towards a Formal Characterization of User Simulation Objectives in Conversational Information Access\\ 
    \bottomrule
    \end{tabularx}
    \caption{List of all lighting talks. Presenters in bold.}
    \label{tab:lighting}    
\end{table}

\section{Summary of Breakout Groups}

\subsection{Group discussion on shared tasks with user simulators}
\label{gd:shared_task}

In this breakout group, we discussed the idea of having a shared task based on user simulators. We envision the general idea of conducting a shared task to which participants submit user simulators instead of systems for the sake of having better insights into the validity of user simulators. Generally, we envision the shared task to be based on a train/validation/test data split of user logs, where participants can instantiate their simulators with training samples and have their fidelity evaluated after submitting them for evaluation, which is based on how well the simulated interactions align with the real ones of the test data.

More general topics that were covered during the breakout group discussions included measuring how well the simulated users fit reality, what kind of data to use (logs, new or existing test collection resources, etc.), what kind of sustainable data artifacts would emerge from such a shared task, the need for annotators and how to spend the annotation budget, and what type of information access systems to use. 

% Possible evaluation measure: CWL/A, “hidden measure”: participants get scores of one measure for calibration, prediction on another unknown measure
% - Counterfactual element in the experiment, e.g., test out a new user interface (page-based layout to infinite scrolls)

The following other ideas and aspects emerged from the discussions. In the context of the anticipated \textit{first calibrate, then predict\/} setting, participants could be provided with user logs and scores of one calibration measure. The final evaluations are then conducted with the help of unknown or hidden measures. This setup would align with the idea of having counterfactual elements in the evaluations, where the final evaluation is conducted in a different setting, with a possibly different underlying user model. Likewise, the counterfactual element could be a different type of user interface. For instance, the training logs could be obtained from a search result interface with pagination to simulate and evaluate user interaction with a result page based on an infinite scrolling design.  

% - can simulators reproduce system rankings/orders?
% - modeling clicks for a given query, take relevant documents and generate snippets with different levels of quality
% - Modeling users with different types of exposure
% - Different levels: content-based simulations (cf. snippets) vs. behavior-based simulations (e.g., action sequences)
% - Next action prediction given a set of previous actions (what kind of action, what kind of modality?)

% - Domain-specific tasks legal, health, e-com.

Other suggestions from the audience highlighted the C/W/L framework~\citep{DBLP:journals/tois/MoffatBST17} and the corresponding evaluation toolkit cwl\_eval~\citep{DBLP:conf/sigir/AzzopardiTM19} regarding existing evaluation methods. Similarly, an evaluation scenario could be based on the Tester approach~\citep{DBLP:conf/sigir/LabhishettyZ21,DBLP:conf/ecir/LabhishettyZ22}, where the relative system performance is known the simulators are evaluated by how well they can reproduce the correct system ranking.

Possible sub-tasks could be aligned with different kinds of simulated user behavior. For example, one task could focus on content-based simulations, i.e., where simulated interactions depend on the contents of interface elements and modalities like snippet text, while the other task could focus on behavior-based simulations at a more abstract level that evaluates interaction sequences from a more general perspective.

Considering the challenge of simulating each user simulation step exactly, another shared task design could be based on providing interaction sequences to participants. Their user simulators would then be used to predict the very next interaction step. This design drastically cuts down complexity but, at the same time, would provide an interesting analysis of what we are currently able to achieve with regard to the prediction of next user interactions.  

In general, it would be quite interesting to have a domain-specific focus for such a shared task. For example, the e-commerce setting or the legal and health domain could introduce an interesting novel direction beyond the still somewhat abstract evaluations of earlier work based on news corpora.

One of the most pressing and overarching question was about how to transfer the user simulation setting into a more modern environment beyond the typical list-based retrieval scenario. Considering the pace of recent advancements in the context of conversational systems and agents, these kinds of technologies offer an excellent basis for having a shared task about user simulators in a modern state-of-the-art setting.

\subsection{Group discussion on simulating users}
\label{gd:simulating_users}

\begin{figure}
    \centering
    \includegraphics[width=0.45\linewidth]{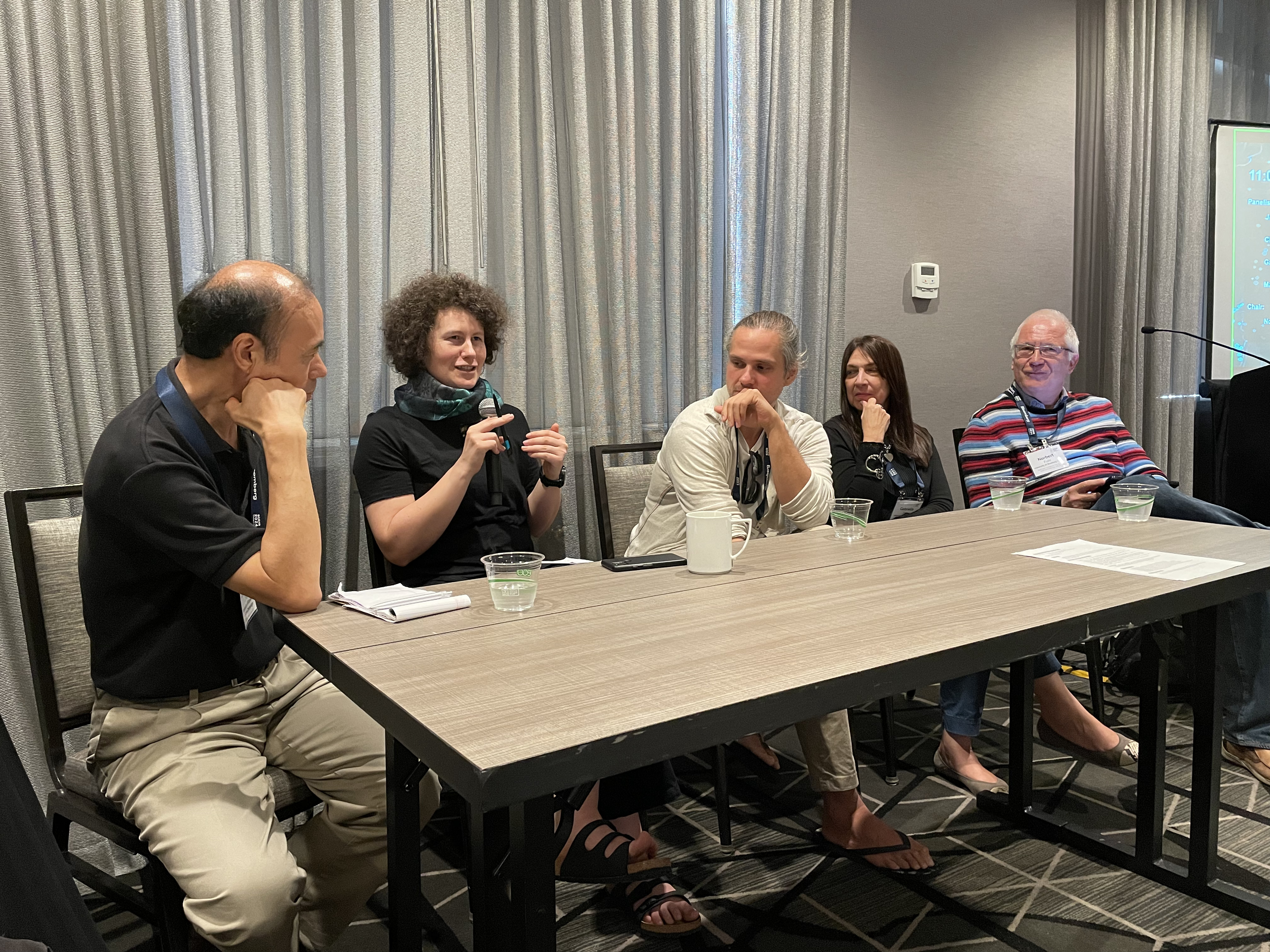}
    \includegraphics[width=0.45\linewidth]{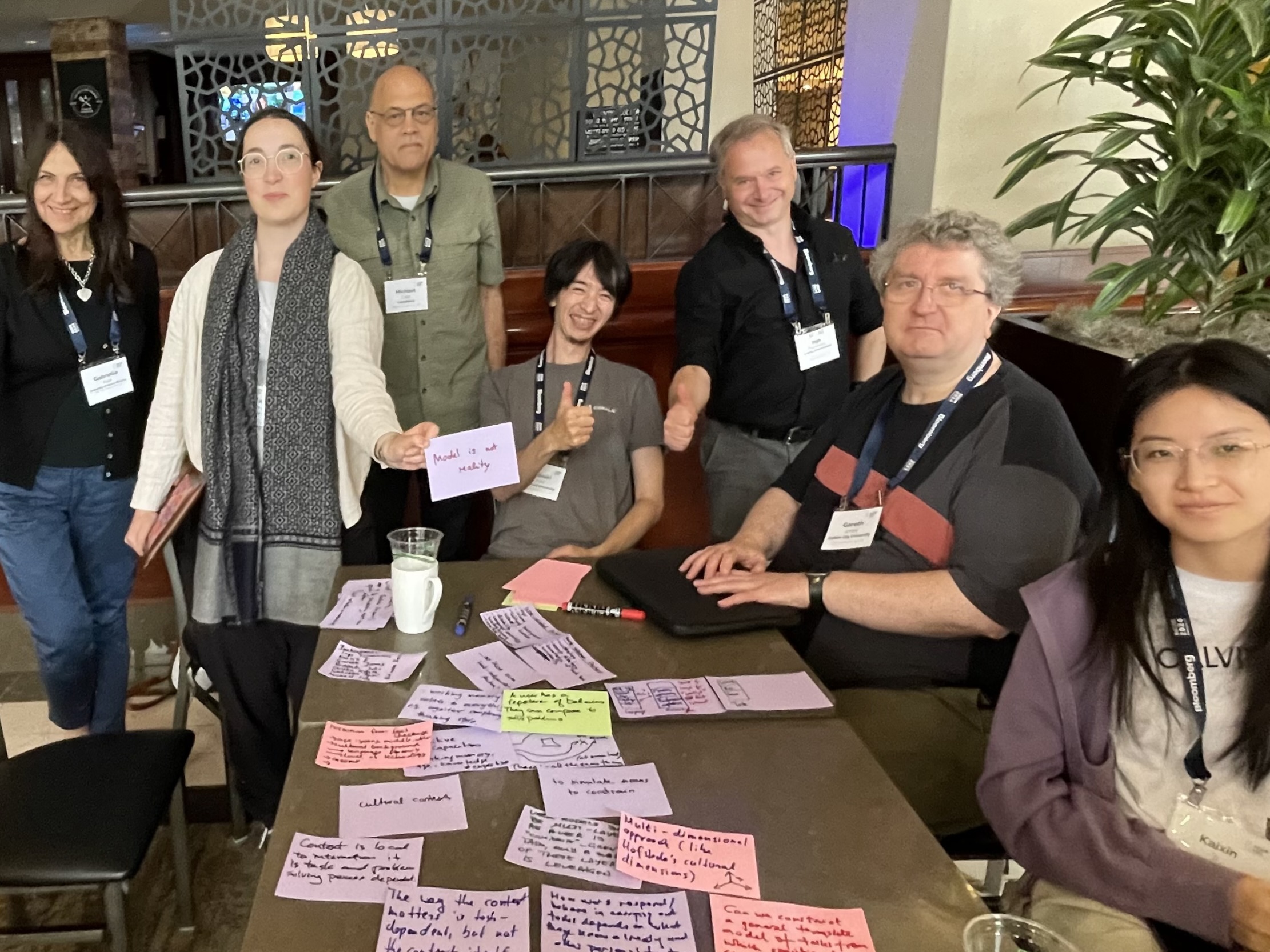}
    \caption{Left: Panel discussion. Right: Breakout group participants.}
    \label{fig:panel_and_breakout}
\end{figure}

In this breakout group (see right side of \autoref{fig:panel_and_breakout}), we examined the question of which user archetypes or personas we should model. We did so by thinking about \textit{factors} of users and/or tasks that would be relevant when trying to represent a user. We considered user-centric factors as those that are independent of a task and do not change when facing a different task.
As \emph{user-centric factors} we mentioned age, income, cultural background, the learning type of a user (e.g., visual), disability, language knowledge and fluency, working memory, level of technology knowledge and cognitive background. 
Contrasting this, we defined \emph{task-centric factors} as those that are dependent on the task and change, if another task is considered for the same simulated user. We noted the vocabulary used in the task, a task's complexity, the search strategy a user is employing and a user's knowledge and interest of the task.
Furthermore, there are more factors influencing user behavior. A user has a repertoire of behaviors they can compose and solve problems with.

When representing users as information seekers, different  \textit{contexts} plays a role as user-centric factors. Local context partially depends on the task and the problem solving process. Global contexts can be cultural, organisational or societal~\cite[Ch.\ 6]{ingwersen_turn_2005}.

Interaction effects appear in each level of user modeling. Interaction should be described in a formal framework, as we may not know the processes that create the interaction effects. When modeling interaction effects mathematically, should the base spaces (e.g.\ tasks) be discrete or continuous? 
%\todo[inline]{base space?}
%\todo[inline]{How could such a formal framework for interaction look like? Fuhr's iPRP for example? Zhai's Card Model? Something else? -- IF}

The question arose if we could construct a general template model of tasks from which specific tasks can be described.

One participant proposed to use reinforcement learning from user profiles while another wondered how we would avoid combinatorial explosion when we consider tasks, user model and context at the same time.

\begin{figure}
    \centering
    \includegraphics[width=0.7\linewidth]{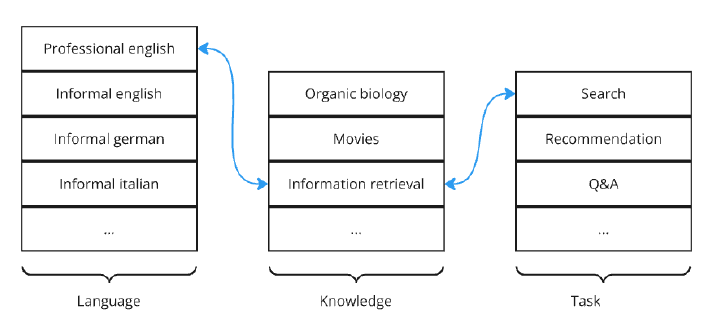}
    \caption{Depiction of the multi-layered factors of a user for the example task of learning about evaluation methodologies for RAG.}
    \label{fig:layers}
\end{figure}

We concluded that to simulate means to constrain and that a user model is not, and cannot be, reality~–- every model tries to approximate reality as closely as possible. A user is a \textit{composit}.
User models in simulation should be multi-dimensional, for example in terms of Hofstede's cultural dimensions~\citep{Hofstede:2010:book}, or (not necessarily mutually exclusive) multi-layered. In the multi-layered representation, each layer is a stack of different levels of skills or knowledge. When given a task, only a subset of these layers is leveraged and within each layer specific skills are selected (see Figure~\ref{fig:layers}). 

As a second smaller topic this breakout group shared some thoughts on \textit{evaluation}. We composed a set of questions to which answers could help mitigate the uncertainty we are currently facing: What does it mean to evaluate the quality of a user profile? Is user profile evaluation the \say{same} as simulator evaluation? One perspective could be to consider them distinct and regard the simulation as a process, e.g., a personalization process. When do we have to consider a user profile's development over time and when could it also be enough to only focus on a static snapshot from a profile? Would it be easier to evaluate the quality of a user profile representing a group or a single user? 

In terms of a quantification of the quality, we talked about the possible use of Fréchet distance. An evaluation metric could be composed from the ability to predict the next user action based on a current state. An extrinsic evaluation of user profiles might be necessary, e.g., in a search task, since how to best use the profile is unclear. 

\section{Summary and Outlook}

The workshop concluded with many inspiring ideas and directions for follow-ups and revealed challenges ahead. Our keynote speakers made it clear that user simulations are highly important for both industry and academia. Likewise, user simulations help better personalize content for users but also allow system evaluations without involving real users in online experiments.

During the panel discussion, user simulations were discussed from different points of view. The panelists highlighted their merits and potentials but also considered limitations of simulated users. Notably, they agreed that user simulations can bridge the gap between offline and online experiments toward a more user-centric evaluation.

Furthermore, the lightning talks gave valuable insights about ongoing work, including research ideas, resources, and (preliminary) results that involve user simulations for various kinds of information access systems and environments.

Our breakout groups mainly targeted the topics of  organizing a shared task with user simulations (cf. \ref{gd:shared_task}) and defining reasonable user archetypes or personas (cf. \ref{gd:simulating_users}). While we did not succeed with our ambitious goal of having a final shared task definition at the very end, our breakout discussions revealed that there is generally a strong interest in running a shared task that builds upon user simulations.

In this regard, we were able to identify major challenges like the overall question of how we can evaluate the validity of simulated users within a shared task setting or what kinds of user archetypes are worth to be considered in this context. 

Most notably, there is a strong interest in running such a task but also many challenges lie ahead. We conclude that there is a need for additional community work and we envision a follow-up event to this workshop for having a more focused and in-depth discussion with experienced shared task organizers but also interested participants.  

\section*{Acknowledgments}

This workshop has been partially funded by the project PLan\_CV. Within the funding programme FH-Personal, the project PLan\_CV (reference number 03FHP109) is funded by the German Federal Ministry of Education and Research (BMBF) and Joint Science Conference (GWK). DAAD gave a travel grant under funding number 57706671.  

\appendix
\section{Authors and Affiliations}
\label{sec:appendix}

\textbf{Workshop organizers:}

\begin{itemize}
    \item Timo Breuer; TH Köln, Cologne, Germany; timobreuer@acm.org
    \item Christin Katharina Kreutz; TH Mittelhessen, Gießen, Germany; ckreutz@acm.org
    \item Norbert Fuhr; University of Duisburg-Essen, Duisburg, Germany; norbert.fuhr@uni-due.de
    \item Krisztian Balog; University of Stavanger, Stavanger, Norway; krisztian.balog@uis.no
    \item Philipp Schaer; TH Köln, Cologne, Germany; philipp.schaer@th-koeln.de
\end{itemize}

\textbf{Other authors:}

 %\author[marcel.gohsen@uni-weimar.de]{Marcel Gohsen}{Bauhaus-Universität Weimar}{Weimar, Germany}
 %   \and
 %   \author[xi.wang@sheffield.ac.uk]{Xi Wang}{University of Sheffield}{Sheffield, UK}
%    \and
%    \author[jueri.keller@th-koeln.de]{Jüri Keller}{TH Köln}{Cologne, Germany}
    %\and
    %\author[saber.zerhoudi@uni-passau.de]{Saber Zerhoudi}{University of Passau}{Passau, Germany}
    %\and
    %\author[nolwenn.m.bernard@uis.no]{Nolwenn Bernard}{University of Stavanger}{Stavanger, Norway}
    %\and    
    %\author[ifrommholz@acm.org]{Ingo Frommholz}{University of Wolverhampton}{Wolverhampton, UK}
    %\and
    %\author[kaixin.ji@student.rmit.edu.au]{Kaixin Ji}{RMIT University}{Melbourne, Australia}
    %\and
    %\author[jiqunliu@ou.edu]{Jiqun Liu}{The University of Oklahoma}{Norman, OK, USA}
%\and
%\author[gabriella.pasi@unimib.it]{Gabriella Pasi}{University of Milano Bicocca}{Milano, Italy}
%    \and
%    \author[j.trippas@rmit.edu.au]{Johanne Trippas}{RMIT University}{Melbourne, Australia}

\begin{itemize}
    \item Nolwenn Bernard; University of Stavanger, Stavanger, Norway; nolwenn.m.bernard@uis.no
    \item Ingo Frommholz; University of Wolverhampton, Wolverhampton, UK; ifrommholz@acm.org
    \item Marcel Gohsen; Bauhaus-Universität Weimar, Weimar, Germany; marcel.gohsen@uni-weimar.de
    \item Kaixin Ji; RMIT University, Melbourne, Australia; kaixin.ji@student.rmit.edu.au
    \item Gareth J. F. Jones; Dublin City University, Dublin, Ireland; Gareth.Jones@dcu.ie
    \item Jüri Keller; TH Köln, Cologne, Germany; jueri.keller@th-koeln.de
    \item Jiqun Liu; The University of Oklahoma, Norman, OK, USA; jiqunliu@ou.edu
    \item Martin Mladenov; Google; mmladenov@google.com
    \item Gabriella Pasi; University of Milano Bicocca, Milano, Italy; gabriella.pasi@unimib.it
    \item Johanne Trippas; RMIT University, Melbourne, Australia; j.trippas@rmit.edu.au
    \item Xi Wang; University of Sheffield, Sheffield, UK; xi.wang@sheffield.ac.uk
    \item Saber Zerhoudi; University of Passau, Passau, Germany; saber.zerhoudi@uni-passau.de
    \item ChengXiang Zhai; University of Illinois at Urbana-Champaign,  Champaign, IL, USA; czhai@illinois.edu

\end{itemize}

\bibliography{sigirforum}

\begin{thebibliography}{27}
\providecommand{\natexlab}[1]{#1}
\providecommand{\url}[1]{\texttt{#1}}
\expandafter\ifx\csname urlstyle\endcsname\relax
  \providecommand{\doi}[1]{doi: #1}\else
  \providecommand{\doi}{doi: \begingroup \urlstyle{rm}\Url}\fi

\bibitem[Aliannejadi et~al.(2024)Aliannejadi, Abbasiantaeb, Chatterjee, Dalton, and Azzopardi]{aliannejadi2024trec}
Mohammad Aliannejadi, Zahra Abbasiantaeb, Shubham Chatterjee, Jeffery Dalton, and Leif Azzopardi.
\newblock Trec ikat 2023: The interactive knowledge assistance track overview, 2024.

\bibitem[Azzopardi et~al.(2019)Azzopardi, Thomas, and Moffat]{DBLP:conf/sigir/AzzopardiTM19}
Leif Azzopardi, Paul Thomas, and Alistair Moffat.
\newblock cwl{\_}eval: An evaluation tool for information retrieval.
\newblock In Benjamin Piwowarski, Max Chevalier, {\'{E}}ric Gaussier, Yoelle Maarek, Jian{-}Yun Nie, and Falk Scholer, editors, \emph{Proceedings of the 42nd International {ACM} {SIGIR} Conference on Research and Development in Information Retrieval, {SIGIR} 2019, Paris, France, July 21-25, 2019}, pages 1321--1324. {ACM}, 2019.
\newblock \doi{10.1145/3331184.3331398}.
\newblock URL \url{https://doi.org/10.1145/3331184.3331398}.

\bibitem[Azzopardi et~al.(2024)Azzopardi, Clarke, Kantor, Mitra, Trippas, and Ren]{azzopardi2024report}
Leif Azzopardi, Charles L.~A. Clarke, Paul Kantor, Bhaskar Mitra, Johanne~R. Trippas, and Zhaochun Ren.
\newblock {Report on The Search Futures Workshop at ECIR 2024}.
\newblock \emph{SIGIR Forum}, 58\penalty0 (1), 2024.

\bibitem[Balog and Zhai(2023)]{DBLP:conf/cikm/BalogZ23}
Krisztian Balog and ChengXiang Zhai.
\newblock Tutorial on user simulation for evaluating information access systems.
\newblock In Ingo Frommholz, Frank Hopfgartner, Mark Lee, Michael Oakes, Mounia Lalmas, Min Zhang, and Rodrygo L.~T. Santos, editors, \emph{Proceedings of the 32nd {ACM} International Conference on Information and Knowledge Management, {CIKM} 2023, Birmingham, United Kingdom, October 21-25, 2023}, pages 5200--5203. {ACM}, 2023.
\newblock \doi{10.1145/3583780.3615296}.
\newblock URL \url{https://doi.org/10.1145/3583780.3615296}.

\bibitem[Balog and Zhai(2024)]{balog2024user}
Krisztian Balog and ChengXiang Zhai.
\newblock User simulation for evaluating information access systems.
\newblock \emph{Foundations and Trends in Information Retrieval}, 18\penalty0 (1-2):\penalty0 1--261, 2024.
\newblock \doi{10.1561/1500000098}.
\newblock URL \url{http://dx.doi.org/10.1561/1500000098}.

\bibitem[Balog et~al.(2021)Balog, Maxwell, Thomas, and Zhang]{DBLP:journals/sigir/BalogMTZ21}
Krisztian Balog, David Maxwell, Paul Thomas, and Shuo Zhang.
\newblock Report on the 1st simulation for information retrieval workshop (sim4ir 2021) at {SIGIR} 2021.
\newblock \emph{{SIGIR} Forum}, 55\penalty0 (2):\penalty0 10:1--10:16, 2021.
\newblock \doi{10.1145/3527546.3527559}.
\newblock URL \url{https://doi.org/10.1145/3527546.3527559}.

\bibitem[Bernard and Balog(2024)]{10.1145/3664190.3672529}
Nolwenn Bernard and Krisztian Balog.
\newblock Towards a formal characterization of user simulation objectives in conversational information access.
\newblock In \emph{Proceedings of the 2024 ACM SIGIR International Conference on Theory of Information Retrieval}, ICTIR '24, page 185–193, New York, NY, USA, 2024. Association for Computing Machinery.
\newblock ISBN 9798400706813.
\newblock \doi{10.1145/3664190.3672529}.
\newblock URL \url{https://doi.org/10.1145/3664190.3672529}.

\bibitem[Chen et~al.(2023)Chen, Liu, and Sakai]{chen2023reference}
Nuo Chen, Jiqun Liu, and Tetsuya Sakai.
\newblock A reference-dependent model for web search evaluation: Understanding and measuring the experience of boundedly rational users.
\newblock In \emph{Proceedings of the ACM Web Conference 2023}, pages 3396--3405, 2023.

\bibitem[Gonzalo et~al.(2009)Gonzalo, Peinado, Clough, and Karlgren]{DBLP:conf/clef/GonzaloPCK09}
Julio Gonzalo, V{\'{\i}}ctor Peinado, Paul~D. Clough, and Jussi Karlgren.
\newblock Overview of iclef 2009: Exploring search behaviour in a multilingual folksonomy environment.
\newblock In Carol Peters, Barbara Caputo, Julio Gonzalo, Gareth J.~F. Jones, Jayashree Kalpathy{-}Cramer, Henning M{\"{u}}ller, and Theodora Tsikrika, editors, \emph{Multilingual Information Access Evaluation {II.} Multimedia Experiments - 10th Workshop of the Cross-Language Evaluation Forum, {CLEF} 2009, Corfu, Greece, September 30 - October 2, 2009, Revised Selected Papers}, volume 6242 of \emph{Lecture Notes in Computer Science}, pages 13--20. Springer, 2009.
\newblock \doi{10.1007/978-3-642-15751-6\_2}.
\newblock URL \url{https://doi.org/10.1007/978-3-642-15751-6\_2}.

\bibitem[Hofstede et~al.(2010)Hofstede, Hofstede, and Minkov]{Hofstede:2010:book}
Geert Hofstede, Gert~Jan Hofstede, and Michael Minkov.
\newblock \emph{Cultures and Organizations: Software of the Mind, Third Edition}.
\newblock McGraw-Hill Professional, 2010.

\bibitem[Hopfgartner et~al.(2019)Hopfgartner, Balog, Lommatzsch, Kelly, Kille, Schuth, and Larson]{hopfgartner_continuous_2019}
Frank Hopfgartner, Krisztian Balog, Andreas Lommatzsch, Liadh Kelly, Benjamin Kille, Anne Schuth, and Martha Larson.
\newblock Continuous {Evaluation} of {Large}-{Scale} {Information} {Access} {Systems}: {A} {Case} for {Living} {Labs}.
\newblock In Nicola Ferro and Carol Peters, editors, \emph{Information {Retrieval} {Evaluation} in a {Changing} {World}}, volume~41, pages 511--543. Springer International Publishing, Cham, 2019.
\newblock ISBN 978-3-030-22947-4 978-3-030-22948-1.
\newblock \doi{10.1007/978-3-030-22948-1_21}.
\newblock URL \url{http://link.springer.com/10.1007/978-3-030-22948-1_21}.
\newblock Series Title: The Information Retrieval Series.

\bibitem[Hsu et~al.(2024)Hsu, Mladenov, Meshi, Pine, Pham, Li, Liang, Polishko, Yang, Scheetz, and Boutilier]{10.1145/3626772.3661358}
Chih-Wei Hsu, Martin Mladenov, Ofer Meshi, James Pine, Hubert Pham, Shane Li, Xujian Liang, Anton Polishko, Li~Yang, Ben Scheetz, and Craig Boutilier.
\newblock Minimizing live experiments in recommender systems: User simulation to evaluate preference elicitation policies.
\newblock In \emph{Proceedings of the 47th International ACM SIGIR Conference on Research and Development in Information Retrieval}, SIGIR '24, page 2925–2929, New York, NY, USA, 2024. Association for Computing Machinery.
\newblock ISBN 9798400704314.
\newblock \doi{10.1145/3626772.3661358}.
\newblock URL \url{https://doi.org/10.1145/3626772.3661358}.

\bibitem[Ingwersen and Järvelin(2005)]{ingwersen_turn_2005}
Peter Ingwersen and Kalervo Järvelin.
\newblock \emph{The turn - integration of information seeking and retrieval in context}.
\newblock Springer, Dordrecht, 2005.
\newblock ISBN 978-1-4020-3851-8.

\bibitem[Jagerman et~al.(2017)Jagerman, Balog, Schaer, Schaible, Tavakolpoursaleh, and de~Rijke]{jagerman-2017-overview}
Rolf Jagerman, Krisztian Balog, Philipp Schaer, Johann Schaible, Narges Tavakolpoursaleh, and Maarten de~Rijke.
\newblock Overview of {TREC} opensearch 2017.
\newblock In Ellen~M. Voorhees and Angela Ellis, editors, \emph{Proceedings of The Twenty-Sixth Text REtrieval Conference, {TREC} 2017, Gaithersburg, Maryland, USA, November 15-17, 2017}, volume 500-324 of \emph{{NIST} Special Publication}. National Institute of Standards and Technology {(NIST)}, 2017.
\newblock URL \url{https://trec.nist.gov/pubs/trec26/papers/Overview-O.pdf}.

\bibitem[Kelly(2009)]{kelly_methods_2009}
Diane Kelly.
\newblock Methods for {Evaluating} {Interactive} {Information} {Retrieval} {Systems} with {Users}.
\newblock \emph{Foundations and Trends in Information Retrieval}, 3\penalty0 (1—2):\penalty0 1--224, 2009.
\newblock ISSN 1554-0669.
\newblock \doi{10.1561/1500000012}.
\newblock URL \url{http://www.nowpublishers.com/product.aspx?product=INR&doi=1500000012}.

\bibitem[Kiesel et~al.(2024)Kiesel, Gohsen, Mirzakhmedova, Hagen, and Stein]{10.1007/978-3-031-56060-6_25}
Johannes Kiesel, Marcel Gohsen, Nailia Mirzakhmedova, Matthias Hagen, and Benno Stein.
\newblock Simulating follow-up questions in conversational search.
\newblock In Nazli Goharian, Nicola Tonellotto, Yulan He, Aldo Lipani, Graham McDonald, Craig Macdonald, and Iadh Ounis, editors, \emph{Advances in Information Retrieval}, pages 382--398, Cham, 2024. Springer Nature Switzerland.
\newblock ISBN 978-3-031-56060-6.

\bibitem[Labhishetty and Zhai(2021)]{DBLP:conf/sigir/LabhishettyZ21}
Sahiti Labhishetty and Chengxiang Zhai.
\newblock An exploration of tester-based evaluation of user simulators for comparing interactive retrieval systems.
\newblock In Fernando Diaz, Chirag Shah, Torsten Suel, Pablo Castells, Rosie Jones, and Tetsuya Sakai, editors, \emph{{SIGIR} '21: The 44th International {ACM} {SIGIR} Conference on Research and Development in Information Retrieval, Virtual Event, Canada, July 11-15, 2021}, pages 1598--1602. {ACM}, 2021.
\newblock \doi{10.1145/3404835.3463091}.
\newblock URL \url{https://doi.org/10.1145/3404835.3463091}.

\bibitem[Labhishetty and Zhai(2022)]{DBLP:conf/ecir/LabhishettyZ22}
Sahiti Labhishetty and ChengXiang Zhai.
\newblock {RATE:} {A} reliability-aware tester-based evaluation framework of user simulators.
\newblock In Matthias Hagen, Suzan Verberne, Craig Macdonald, Christin Seifert, Krisztian Balog, Kjetil N{\o}rv{\aa}g, and Vinay Setty, editors, \emph{Advances in Information Retrieval - 44th European Conference on {IR} Research, {ECIR} 2022, Stavanger, Norway, April 10-14, 2022, Proceedings, Part {I}}, volume 13185 of \emph{Lecture Notes in Computer Science}, pages 336--350. Springer, 2022.
\newblock \doi{10.1007/978-3-030-99736-6\_23}.
\newblock URL \url{https://doi.org/10.1007/978-3-030-99736-6\_23}.

\bibitem[Liu and Shah(2019)]{liu2019interactive}
Jiqun Liu and Chirag Shah.
\newblock \emph{Interactive IR user study design, evaluation, and reporting}.
\newblock Morgan \& Claypool Publishers, 2019.

\bibitem[Moffat et~al.(2017)Moffat, Bailey, Scholer, and Thomas]{DBLP:journals/tois/MoffatBST17}
Alistair Moffat, Peter Bailey, Falk Scholer, and Paul Thomas.
\newblock Incorporating user expectations and behavior into the measurement of search effectiveness.
\newblock \emph{{ACM} Trans. Inf. Syst.}, 35\penalty0 (3):\penalty0 24:1--24:38, 2017.
\newblock \doi{10.1145/3052768}.
\newblock URL \url{https://doi.org/10.1145/3052768}.

\bibitem[Penha et~al.(2022)Penha, C{\^{a}}mara, and Hauff]{DBLP:conf/ecir/PenhaCH22}
Gustavo Penha, Arthur C{\^{a}}mara, and Claudia Hauff.
\newblock Evaluating the robustness of retrieval pipelines with query variation generators.
\newblock In Matthias Hagen, Suzan Verberne, Craig Macdonald, Christin Seifert, Krisztian Balog, Kjetil N{\o}rv{\aa}g, and Vinay Setty, editors, \emph{Advances in Information Retrieval - 44th European Conference on {IR} Research, {ECIR} 2022, Stavanger, Norway, April 10-14, 2022, Proceedings, Part {I}}, volume 13185 of \emph{Lecture Notes in Computer Science}, pages 397--412. Springer, 2022.
\newblock \doi{10.1007/978-3-030-99736-6\_27}.
\newblock URL \url{https://doi.org/10.1007/978-3-030-99736-6\_27}.

\bibitem[Sadiri~Javadi et~al.(2023)Sadiri~Javadi, Potthast, and Flek]{sadiri-javadi-etal-2023-opinionconv}
Vahid Sadiri~Javadi, Martin Potthast, and Lucie Flek.
\newblock {O}pinion{C}onv: Conversational product search with grounded opinions.
\newblock In Svetlana Stoyanchev, Shafiq Joty, David Schlangen, Ondrej Dusek, Casey Kennington, and Malihe Alikhani, editors, \emph{Proceedings of the 24th Annual Meeting of the Special Interest Group on Discourse and Dialogue}, pages 66--76, Prague, Czechia, September 2023. Association for Computational Linguistics.
\newblock \doi{10.18653/v1/2023.sigdial-1.6}.
\newblock URL \url{https://aclanthology.org/2023.sigdial-1.6}.

\bibitem[Schaer et~al.(2021)Schaer, Breuer, Castro, Wolff, Schaible, and Tavakolpoursaleh]{schaer-2021-overview}
Philipp Schaer, Timo Breuer, Leyla~Jael Castro, Benjamin Wolff, Johann Schaible, and Narges Tavakolpoursaleh.
\newblock Overview of lilas 2021 - living labs for academic search (extended overview).
\newblock In Guglielmo Faggioli, Nicola Ferro, Alexis Joly, Maria Maistro, and Florina Piroi, editors, \emph{Proceedings of the Working Notes of {CLEF} 2021 - Conference and Labs of the Evaluation Forum, Bucharest, Romania, September 21st - to - 24th, 2021}, volume 2936 of \emph{{CEUR} Workshop Proceedings}, pages 1668--1699. CEUR-WS.org, 2021.
\newblock URL \url{https://ceur-ws.org/Vol-2936/paper-143.pdf}.

\bibitem[Schaer et~al.(2024)Schaer, Kreutz, Balog, Breuer, and Fuhr]{Schaer2024}
Philipp Schaer, Christin~Katharina Kreutz, Krisztian Balog, Timo Breuer, and Norbert Fuhr.
\newblock Sigir 2024 workshop on simulations for information access (sim4ia 2024).
\newblock In \emph{Proceedings of the 47th International ACM SIGIR Conference on Research and Development in Information Retrieval}, SIGIR '24, page 3058–3061, New York, NY, USA, 7 2024. Association for Computing Machinery.
\newblock ISBN 9798400704314.
\newblock \doi{10.1145/3626772.3657991}.
\newblock URL \url{https://doi.org/10.1145/3626772.3657991}.

\bibitem[Wang et~al.(2024)Wang, Sen, Li, and Yilmaz]{10.1007/978-3-031-56027-9_10}
Xi~Wang, Procheta Sen, Ruizhe Li, and Emine Yilmaz.
\newblock Simulated task oriented dialogues for developing versatile conversational agents.
\newblock In \emph{Advances in Information Retrieval: 46th European Conference on Information Retrieval, ECIR 2024, Glasgow, UK, March 24–28, 2024, Proceedings, Part I}, page 157–172, Berlin, Heidelberg, 2024. Springer-Verlag.
\newblock ISBN 978-3-031-56026-2.
\newblock \doi{10.1007/978-3-031-56027-9_10}.
\newblock URL \url{https://doi.org/10.1007/978-3-031-56027-9_10}.

\bibitem[Zerhoudi et~al.(2022)Zerhoudi, G\"{u}nther, Plassmeier, Borst, Seifert, Hagen, and Granitzer]{10.1145/3511808.3557711}
Saber Zerhoudi, Sebastian G\"{u}nther, Kim Plassmeier, Timo Borst, Christin Seifert, Matthias Hagen, and Michael Granitzer.
\newblock The simiir 2.0 framework: User types, markov model-based interaction simulation, and advanced query generation.
\newblock In \emph{Proceedings of the 31st ACM International Conference on Information \& Knowledge Management}, CIKM '22, page 4661–4666, New York, NY, USA, 2022. Association for Computing Machinery.
\newblock ISBN 9781450392365.
\newblock \doi{10.1145/3511808.3557711}.
\newblock URL \url{https://doi.org/10.1145/3511808.3557711}.

\bibitem[Zhang et~al.(2024)Zhang, Wang, Gong, Lin, and Mao]{DBLP:conf/sigir/ZhangWGLM24}
Erhan Zhang, Xingzhu Wang, Peiyuan Gong, Yankai Lin, and Jiaxin Mao.
\newblock Usimagent: Large language models for simulating search users.
\newblock In Grace~Hui Yang, Hongning Wang, Sam Han, Claudia Hauff, Guido Zuccon, and Yi~Zhang, editors, \emph{Proceedings of the 47th International {ACM} {SIGIR} Conference on Research and Development in Information Retrieval, {SIGIR} 2024, Washington DC, USA, July 14-18, 2024}, pages 2687--2692. {ACM}, 2024.
\newblock \doi{10.1145/3626772.3657963}.
\newblock URL \url{https://doi.org/10.1145/3626772.3657963}.

\end{thebibliography}

\end{document}